\documentclass[pre,showkeys,showpacs]{revtex4}
\usepackage{graphicx}
\begin{document}
\title{On the exact number of possibilities for cutting and reconnecting the
tour of a traveling salesman with Lin-$k$-Opts}

\author{G\"unther Stattenberger}
\email{gstattenberger@fernfachhochschule.ch}
\homepage{http://www.iam.unibe.ch/~stattenb}
\affiliation{Fernfachhochschule Schweiz, \"Uberlandstr.\ 12,
Postfach 689, CH-3900 Brig, Switzerland}

\author{Markus Dankesreiter}
\email{Markus.Dankesreiter@web.de}
\affiliation{Ed.-M\"uhlbauer-Weg 14, D-93051 Regensburg, Germany}

\author{Florian Baumgartner}
\email{florian.baumgartner@swisscom.com}
\affiliation{Swisscom Innovations, Ostermundigenstr.\ 93,
CH-3006 Berne, Switzerland}

\author{Johannes J.\ Schneider}
\email{schneidj@uni-mainz.de}
\homepage{http://www.staff.uni-mainz.de/schneidj}
\affiliation{Institute of Physics, Johannes Gutenberg University of Mainz,
Staudinger Weg 7, D-55099 Mainz, Germany}

\date{\today}

\begin{abstract}
When trying to find approximate solutions for the Traveling
Salesman Problem with heuristic optimization algorithms, small moves called
Lin-$k$-Opts are often used. In our paper, we provide exact
formulas for the numbers of possible tours into which a randomly
chosen tour can be changed with a Lin-$k$-Opt.
\end{abstract}
\maketitle

\section{Introduction: The Traveling Salesman Problem}
Due to the simplicity of its formulation and the complexity
of its exact solution, the traveling salesman problem (TSP) has
been studied for a very long time \cite{geschichte} and has
drawn great attention from various fields, such as
applied mathematics, computational physics, and
operations research. The traveling salesman faces the
problem to find the shortest closed tour through a given
set of nodes, touching each of the $N$ nodes exactly once and returning
to the starting node at the end \cite{geschichte,Lawler}.
Hereby the salesman knows
the distances $d(i,j)$ between all pairs
$(i,j)$ of nodes, which are usually given
as some constant non-negative values, either in units of length
or of time. The costs ${\cal H}$ of a configuration
$\sigma$ are therefore given
as the sum of the distances of the used edges. If denoting
a configuration as a permutation $\sigma$ of the numbers
$\{1,\dots,N\}$, the costs can be written as
\begin{equation}
{\cal H}(\sigma)=d(\sigma(N),\sigma(1))+\sum_{i=1}^{N-1}
d(\sigma(i),\sigma(i+1)) .
\end{equation}
A TSP instance
is called symmetric if $d(i,j)=d(j,i)$ for all pairs $(i,j)$ of
nodes. For a symmetric TSP, the costs for going through the
tour in a clockwise direction are the same as going through in
an anticlockwise direction. Thus, these two tours are to be
considered as identical.

\begin{figure} \centering
\includegraphics[width=\columnwidth]{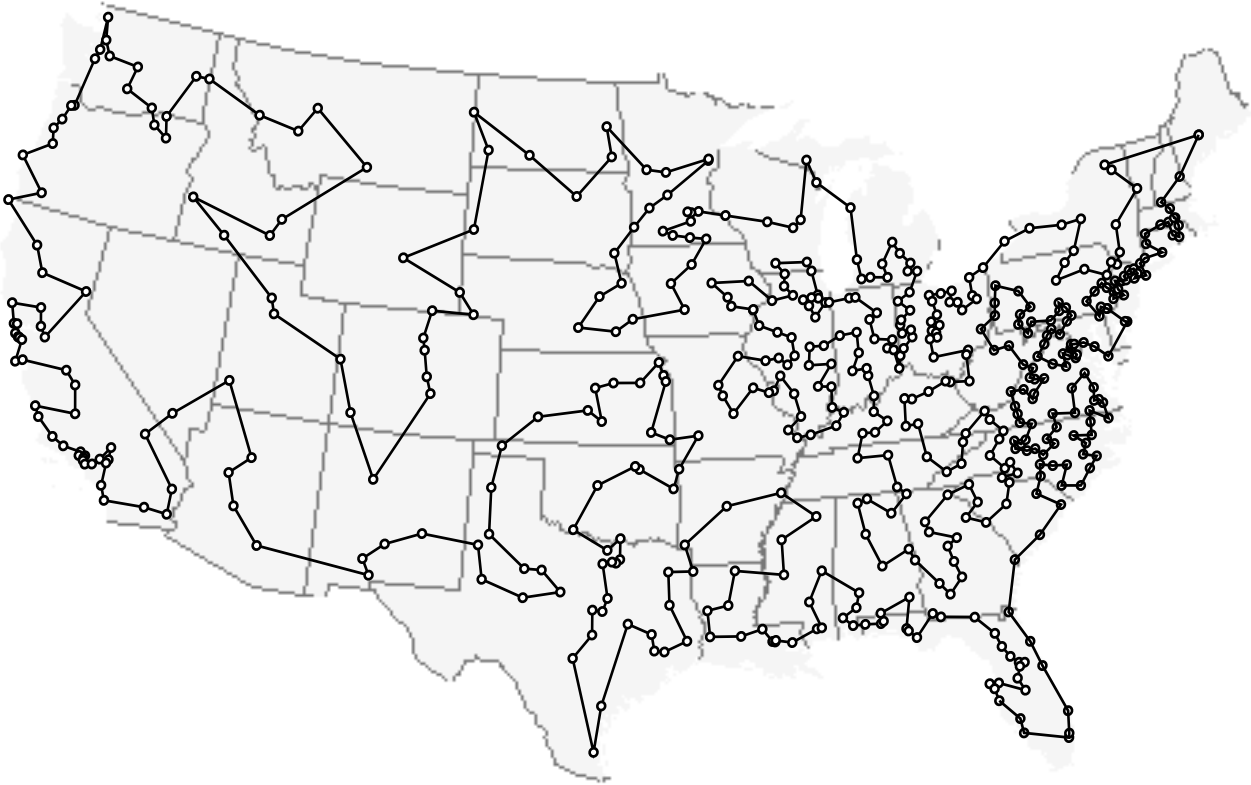}
\caption{The optimum solution of the ATT532 benchmark
TSP instance containing the 532 AT\&T switch locations
in the United States of America: this benchmark instance
is part of Reinelt's benchmark library TSPLIB95 \cite{TSPLIB95}.
The optimum solution was created with the Searching for
Backbones algorithm \cite{sfb,sfb2}, making use of the Lin-2-Opt and the
Lin-3-Opt, which are shown in Figs.\ \ref{fig:l2o} and \ref{fig:l3o}.}
\label{fig:usa}
\end{figure}

As the time for determining the optimum solution of
a proposed TSP instance grows exponentially with the system size,
the number $N$ of nodes, a large variety of heuristics has been developed in
order to solve this problem approximately. Besides the application of
several different construction heuristics \cite{Reinelt}, which
were either specifically designed for the TSP or altered in order
to enable their application to the TSP, the TSP has been tackled with
various general-purpose improvement heuristics, like
Simulated Annealing \cite{Kirk} and related algorithms
such as Threshold Accepting \cite{TA,Pablo}, the Great
Deluge Algorithm \cite{GDA,Sintflut,Dueckbuch}, algorithms
based on the Tsallis statistics \cite{Penna},
Simulated and Parallel Tempering (methods described in
\cite{MarinariParisi,ParallelTempering2,ParallelTempering1,
ParallelTempering1a,ParallelTempering1b}),
and Search Space Smoothing \cite{JunGu,SSS,SSSKonkurrenz}.
Furthermore Genetic Algorithms
\cite{Schoeneburg,Goldbergkommune,Holland},
Tabu Search \cite{Tabu1} and Scatter Search \cite{Tabu1,Tabu2},
and even Ant Colony Optimization \cite{aco}, Particle
Swarm Optimization \cite{psofull,pso,psobook}, and other
biologically motivated algorithms \cite{Antlion} have been
applied to the TSP. The quality of these algorithms is
compared by creating solutions for
benchmark instances, one of which is shown in
Fig.\ \ref{fig:usa}.

Most of these improvement heuristics apply a series of
so-called small moves to a current configuration or
a set of configurations. In this context, the move being
small means that it does not change a configuration very
much, such that usually the cost of the new tentative configuration which
is to be accepted or rejected according to the acceptance
criterion of the improvement heuristics does not differ
very much from the cost of the current configuration. 
This method of using only small moves is called the Local Search
approach, as the small moves lead to tentative new configurations,
which are close to the previous configuration according to some
metric like the Hamming distance for the TSP: the Hamming distance
between two tours is given by the number of different edges.

\section{The Smallest Moves}
\subsection{The Exchange}
\begin{figure}
\begin{center}
\parbox{\columnwidth}{
\includegraphics[width=.49\columnwidth]{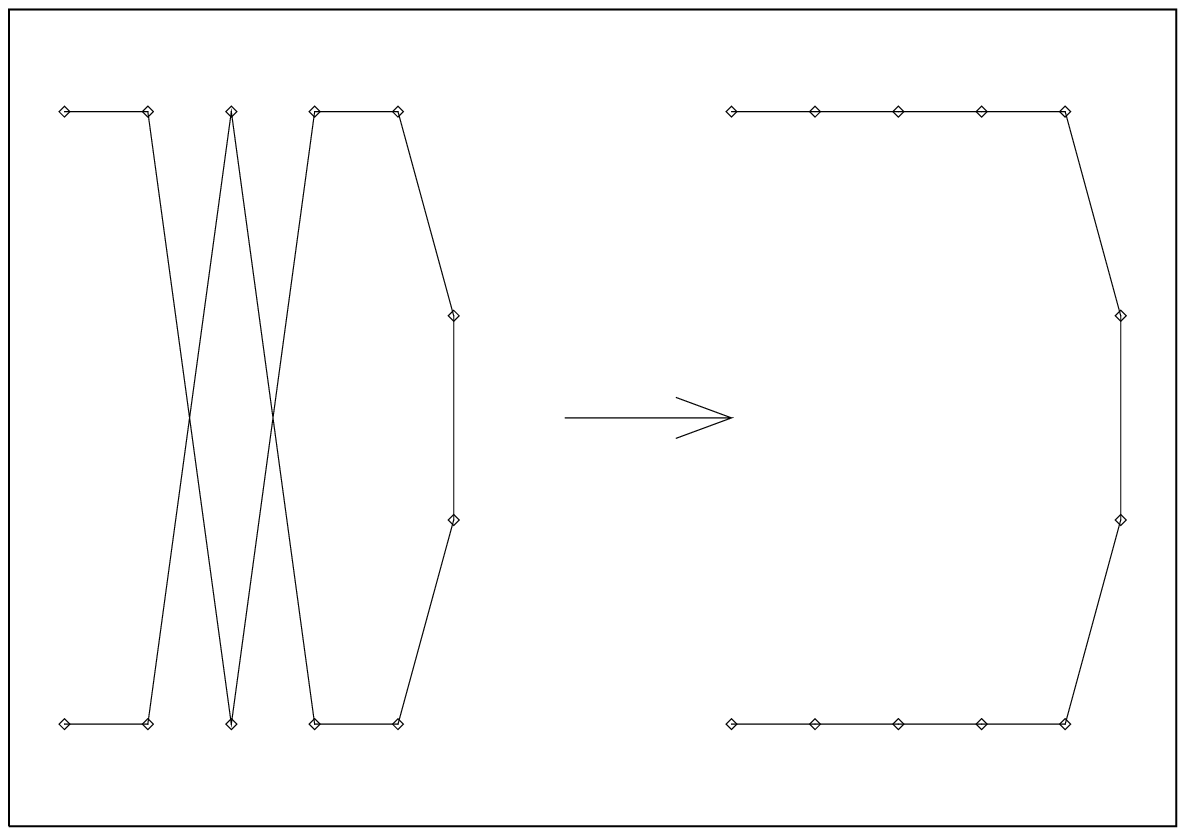}
\hfill
\includegraphics[width=.49\columnwidth]{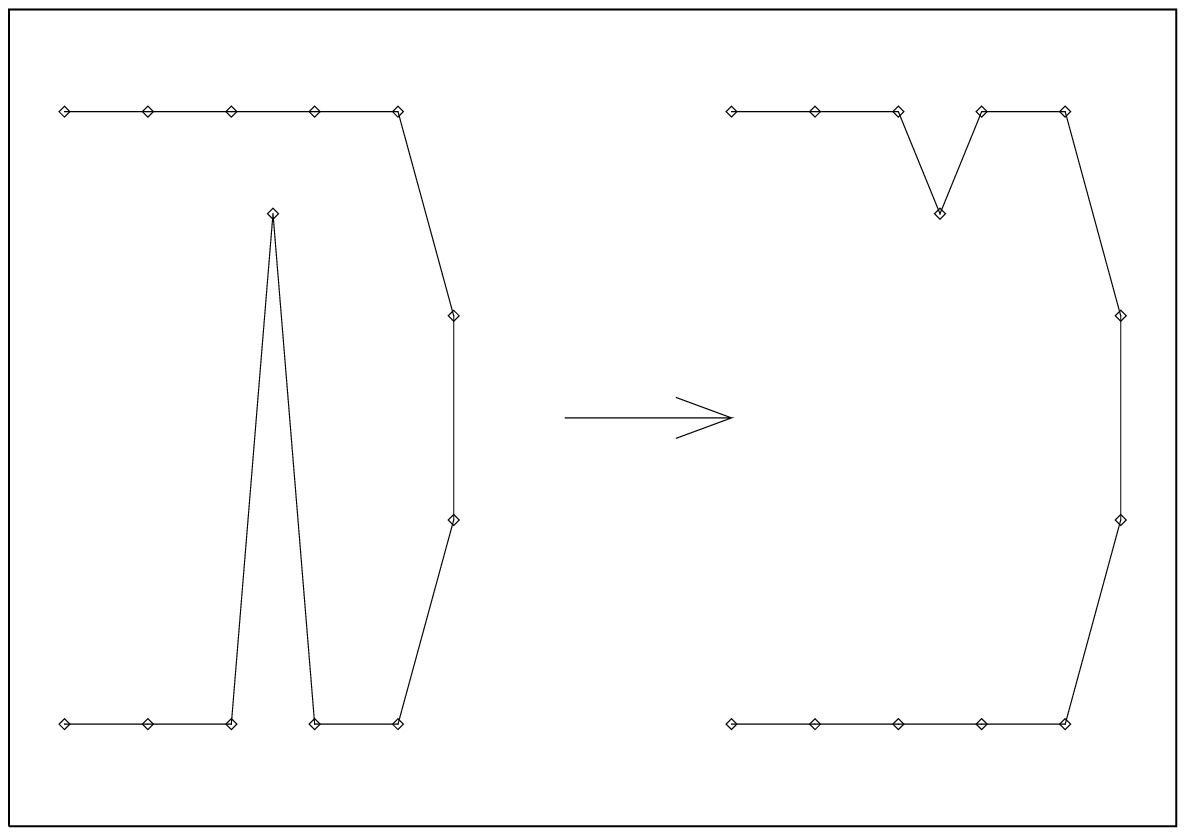}}
\caption{The Exchange (left) and the Node Insertion Move (right)}
\label{fig:excnim}
\end{center}
\end{figure}

One move which does not change a configuration very much is the
Exchange (EXC), which is sometimes also called Swap and which
is shown in Fig.\ \ref{fig:excnim}. The Exchange exchanges two
randomly selected nodes in the tour. Thus, from a proposed configuration,
$N \times (N-1) /2$ other configurations can be reached, such
that the neighborhood of a configuration generated by this move
has a size of order ${\cal O}(N^2)$.

\subsection{The Node Insertion Move}
Another small move is the Node Insertion Move (NIM), which is also
called Jump. It is also shown in Fig.\ \ref{fig:excnim}. The Node Insertion
Move randomly selects a node and an edge. It removes the randomly
chosen node from its original position and places it between the end
points of the randomly selected edge, which is cut for this purpose.
The neighborhood size generated by this move is $N \times (N-2)$
and thus also of order ${\cal O}(N^2)$.

\subsection{The Lin-2-Opt}
\begin{figure}
\begin{center}
\includegraphics[width=.49\columnwidth]{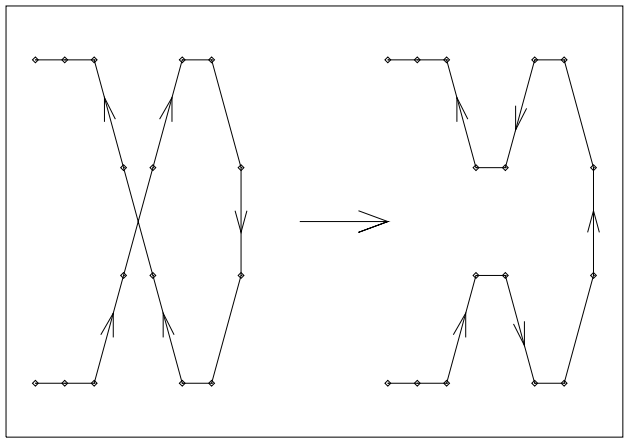}

\caption{The Lin-2-Opt}
\label{fig:l2o}
\end{center}
\end{figure}

Lin introduced a further small move, which is called Lin-2-Opt
(L2O) \cite{Lin,LinK}: as shown in Fig.\ \ref{fig:l2o},
it cuts two edges of the tour, turns
the direction of one of the two partial sequences around, and
reconnects these two sequences in a new way. For symmetric TSP
instances,
only the two removed edges and the two added edges have to be
considered when calculating the cost difference created by this
move. For these symmetric TSPs, it plays no role which of the
two partial sequences is turned around when performing the move,
due to the identical cost function value for moving through
clockwisely or anticlockwisely.
In the symmetric case, on which we will concentrate throughout this
paper, the move creates a neighborhood of size $N \times (N-3)/2$
and thus of order ${\cal O}(N^2)$. Please note that
a move cutting two edges after neighboring nodes does not lead
to a new configuration, such that the neighborhood size is not
$N \times (N-1)/2$, a false value which is sometimes found in the
literature.

The Lin-2-Opt turned out to provide better results for the
symmetric Traveling Salesman Problem than the Exchange.
The reason for this quality difference was explained
analytically by Stadler and Schnabl.
In their paper \cite{Stadler}, they basically found out
that the results are the better the less edges are cut:
the Lin-2-Opt cuts only two edges whereas the Exchange cuts
four. But they also reported results that the Lin-3-Opt
cutting three edges leads to an even better quality of
the solutions than the Lin-2-Opt, what contradicted their
results at first sight, but they
explained this finding with the larger neighborhood size
of the Lin-3-Opt.

\section{The Lin-3-Opt}
\begin{figure}
\begin{center}
\parbox{\columnwidth}{
\includegraphics[width=.49\columnwidth]{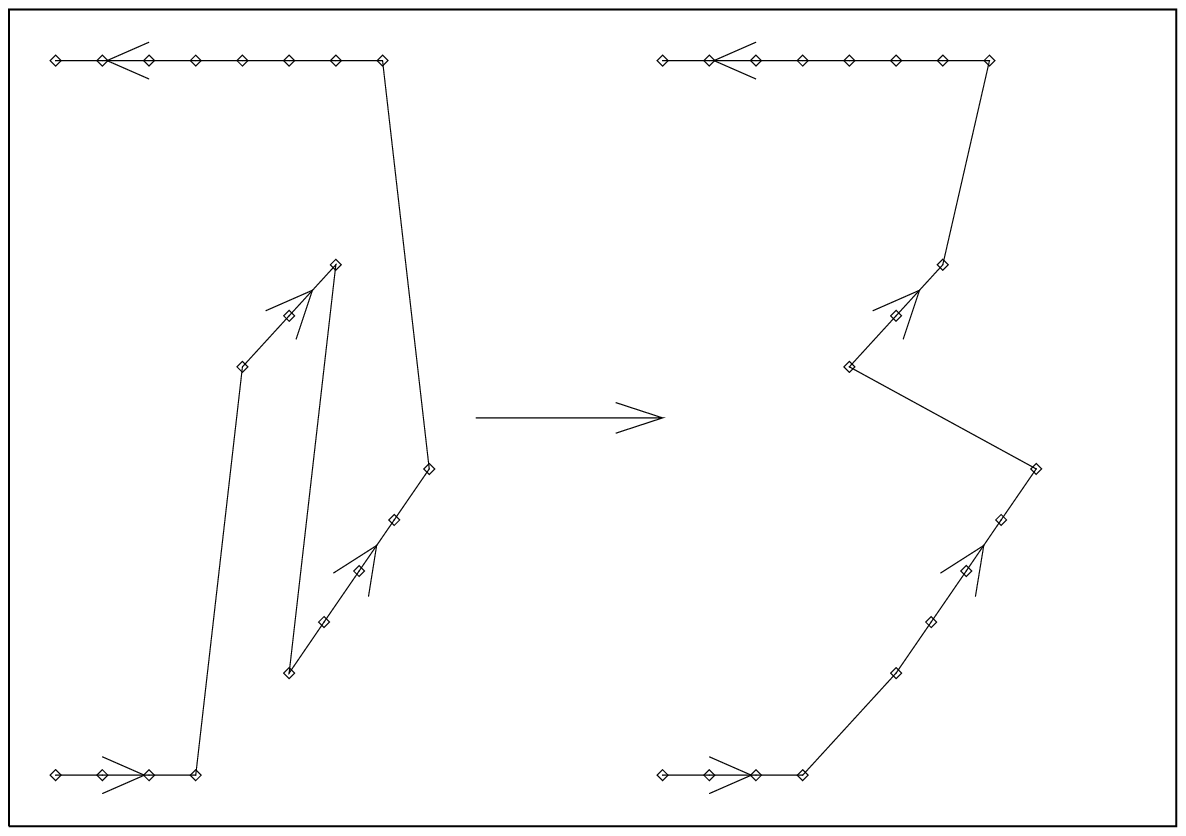}
\hfill
\includegraphics[width=.49\columnwidth]{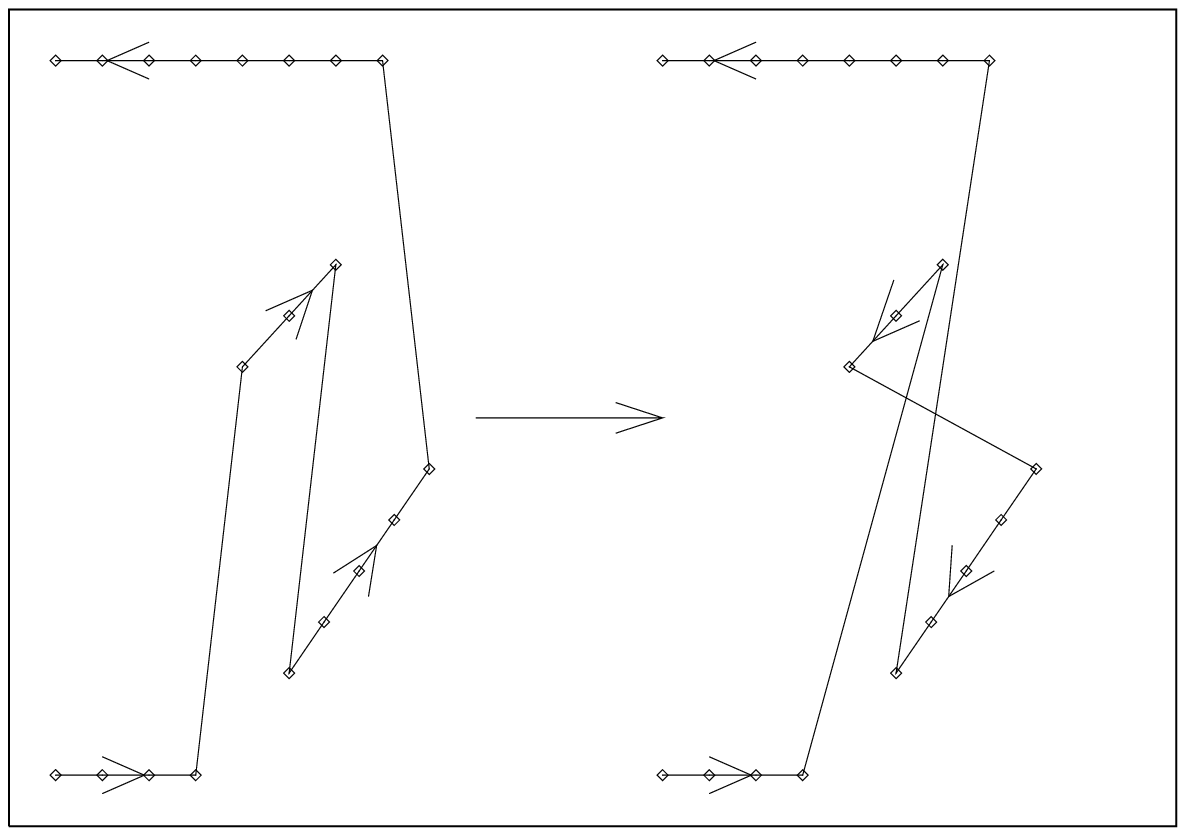}}
\parbox{\columnwidth}{
\includegraphics[width=.49\columnwidth]{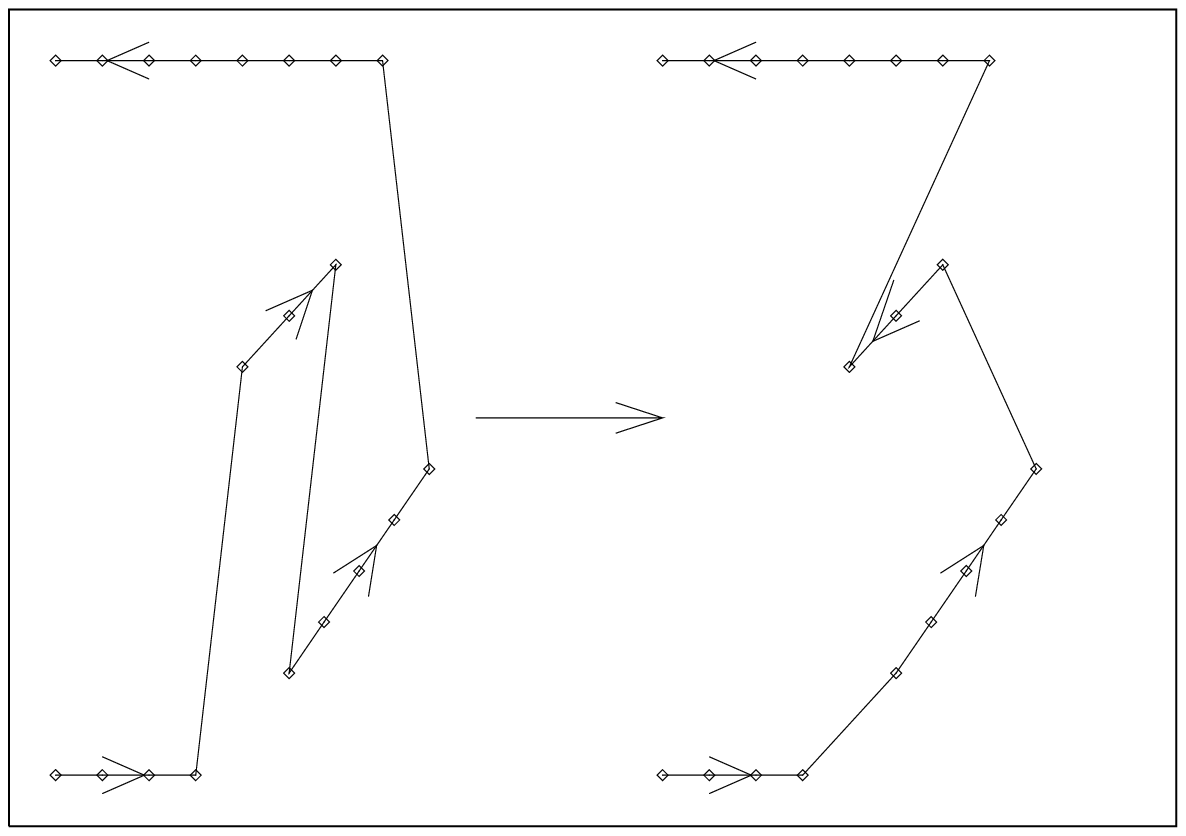}
\hfill
\includegraphics[width=.49\columnwidth]{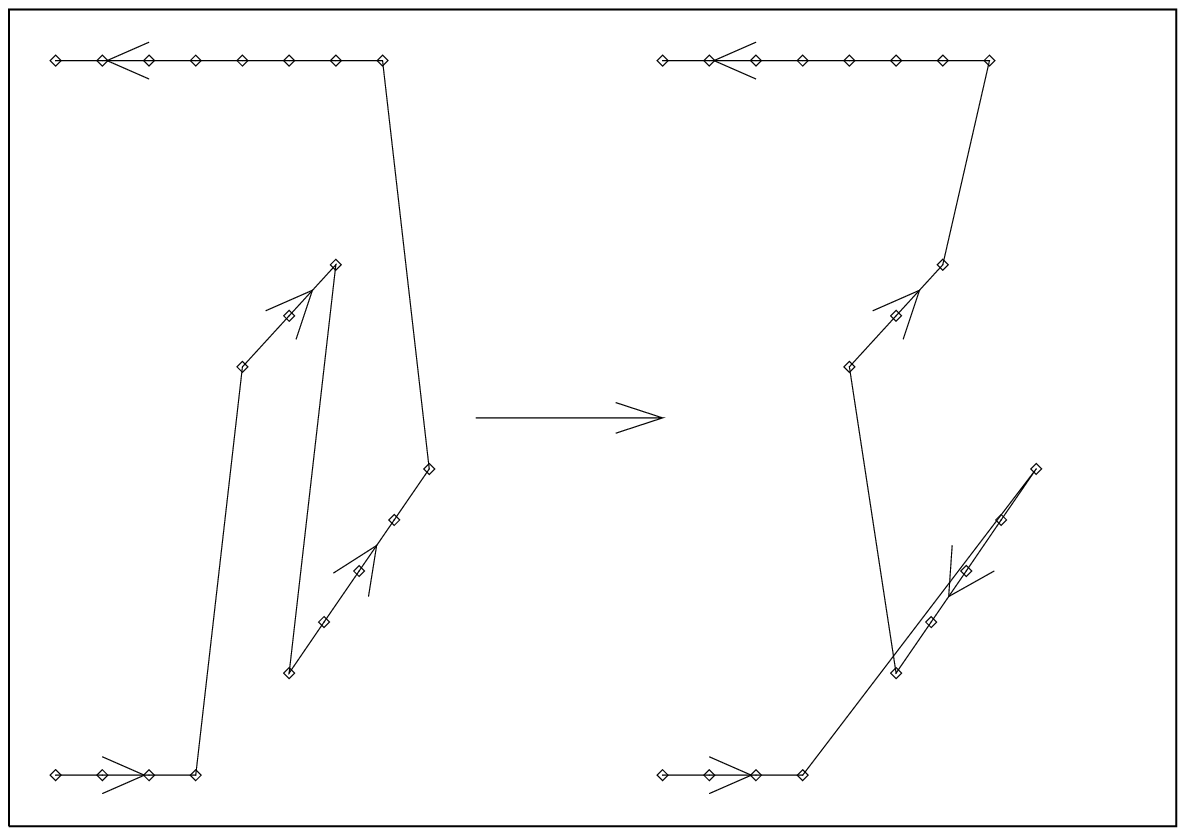}}
\caption{The four possibilities how to change a tour
with a Lin-3-Opt}
\label{fig:l3o}
\end{center}
\end{figure}

The next larger move to the Lin-2-Opt is the Lin-3-Opt (L3O):
the Lin-3-Opt removes three edges of the tour and reconnects
the three partial sequences to some new closed tour. In contrast
to the smallest moves for which there is only one possibility
to create a new tour, there are four possibilities in the
case of the symmetric TSP how to create
a new tour with three new edges with the Lin-3-Opt if each of
the partial sequences contains at least two nodes. These four
possibilities are shown in Fig.\ \ref{fig:l3o}. Please note
that we count only the number of ``true'' possibilities here,
i.e., only those cases in which the tour contains three edges
which were not part formerly in the tour, as otherwise the
move would e.g.\ only be a Lin-2-Opt. If one of the
partial sequences contains only one node and the other two at
least two nodes each, then only one possibility for a
``true'' Lin-3-Opt remains. If even two of the three partial
sequences do only contain one node, then there is no
possibility left to reconnect the three sequences without
reusing at least one of the edges which was cut.
Analogously, there is one possibility for the Lin-2-Opt,
if both partial sequences contain at least two nodes each,
otherwise there is no possibility.

If looking closely at the four variants of the Lin-3-Opt
in Fig.\ \ref{fig:l3o}, one finds that the resulting
configurations could also be generated by a sequence of
Lin-2-Opts: for the upper left variant in Fig.\ \ref{fig:l3o},
three Lin-2-Opts would be needed, whereas only two
Lin-2-Opts would be sufficient for the other three variants.
Thus, one might ask whether the Lin-3-Opt is necessary as
a move as a few Lin-2-Opts could do the same job. However,
due to the acceptance criteria of the improvement heuristics,
it might be that at least one of the Lin-2-Opts would be rejected
whereas the combined Lin-3-Opt move could be accepted.
Thus, it is often advantageous also to implement these
next-higher-order moves in order to overcome the barriers
in the energy landscape of the small moves.

Now the question arises how large the neighborhood size of a
Lin-3-Opt is. Of course, it has to be of order
${\cal O}(N^3)$, as three edges to be removed are
randomly selected out of $N$ edges.
However, for the calculation of the exact number
of possibilities one has to distinguish between the case in
which all partial sequences contain at least two nodes each
and the case in which exactly one partial sequence contains
only one node.

Please note that the Node Insertion Move, which was introduced
earlier, is the special case of the Lin-3-Opt in which one of
the partial sequences only contains one node. But in the
special case that one of
the two next nearest edges to the randomly chosen node is
selected, the NIM corresponds to a Lin-2-Opt. As the number of
cut edges of the NIM is 3, such that this move is basically a Lin-3-Opt,
but as the neighborhood size of this move is of order
${\cal O}(N^2)$, this move is also sometimes called Lin-2.5-Opt.

\section{Higher-order Moves}
One can go on to even higher-order Lin-$k$-Opts: the Lin-4-Opt
cuts four edges of the tour and reconnects the four created
partial sequences in a new way. If every partial sequence contains
at least two nodes, there are 25 possibilities for a true
Lin-4-Opt to reconnect the partial sequences to a closed
feasible tour. The neighborhood size of this move is of order
${\cal O}(N^4)$.

The Exchange, which was also introduced earlier, is usually
a special case of a Lin-4-Opt. Only if the two nodes which
are to be exchanged are direct neighbors of each other or
if there is exactly one node between them, then the move
is equivalent to a Lin-2-Opt.

One can increase the number $k$ of deleted
edges further and further. However, by doing so, one gradually
loses the advantage of the Local Search approach, in which, due
to the similarity of the configurations, their cost values
do not differ much. In the extreme,
the Lin-$N$-Opt would lead to a randomly chosen new configuration,
the cost value of which is not related to the cost value of the
previous configuration at all. Moving away from the Local Search
approach, the probability for getting an acceptable new
configuration among the many more neighboring configurations
with cost values in a much larger interval
strongly decreases due to the finite available computing time.
When using the Local Search approach, it turns out that
using the smallest possible move only is only optimal
in the case of very short computing times. With increasing
computing time, a well chosen combination of the smallest moves
and their next larger variants becomes optimal. Here the
optimization run has more time to search through a larger
neighborhood. The next larger moves enable the system to
overcome barriers in the energy landscape formed by the small
moves only. Of course, one can extend this approach and
also include moves with the next larger $k$ and spend even
more computing time.

However, for some difficult optimization problems,
an approach based on small moves and their next larger
variants is not sufficient \cite{Schrimpf}. There indeed large
moves have to be used. A successful approach here are the
Ruin \& Recreate moves, which destroy a configuration to
some extent and rebuild it according to a given rule set.
They work in
a different way than the small moves, which completely randomly select
a way to change the configuration. In contrast, the Ruin \& Recreate
moves contain constructive elements in order to result in good
configurations. Also for problems like the TSP, for which small
moves basically work, well designed Ruin \& Recreate moves
are superior to the small moves \cite{Schrimpf}.
However, the development of excellent Ruin \& Recreate moves
is rather difficult, it is indeed an optimization problem itself,
whereas the application of the Local Search approach, which simply
intends to ``change the configuration a little bit'', is rather
straightforward and also usually quite successful in producing
good solutions, such that it is mostly used.

Sometimes one needs to know the exact size of the neighborhood generated
by the implemented moves for relating it to the available
computing time or for tuning an optimization algorithm like
Tabu Search. Naturally, one is aware of the neighborhood size
of a Lin-$k$-Opt being of order ${\cal O}(N^k)$. The aim of
this paper is to provide exact numbers for the neighborhood size.
For deriving the neighborhood size of the
Lin-3-Opt and of even larger Lin-$k$-Opts, we will start with
the determination of the number of possibilities for reconnecting
partial sequences to a complete tour with a true
Lin-$k$-Opt in Sec.\ \ref{verbinden}. There we will find
laws how many possibilities exist depending on the number of
partial sequences containing only one node and their
spatial neighborhood relation
to each other. Having this distinction at hand, we will calculate
the corresponding numbers of possibilities for cutting
the tour in Sec.\ \ref{trennen}.

\section{Number of possibilities for reconnecting the tour
with a Lin-$k$-Opt \label{verbinden}}
\subsection{Special case}
\subsubsection{Number of overall possibilities}
For the calculation of the number of possibilities for
reconnecting the tour, we want to start out with the
special case that each partial sequence contains
at least two nodes. A Lin-$k$-Opt cuts the tour into
$k$ partial sequences. The overall number of possibilities
to reconnect them to a closed tour containing all nodes
can be obtained when imagining the following scenario:
one randomly selects one of the partial sequences and fixes
its direction. (This has to be done for the symmetric TSP for
which a tour and its mirror tour are degenerate)
This first partial sequence serves as a starting sequence
for the new tour to be constructed.
Then one randomly selects one out of the $k-1$ remaining
partial sequences and adds it to the already existing
partial tour. There are two possible ways of adding it,
one for each direction of the partial sequence. Thus, one
gets a new system with only $k-1$ partial sequences. The number
of possibilities to construct a new feasible tour is thus
given as
\begin{equation}
P(k)=2 (k-1) P(k-1) .
\end{equation}
This recursive formula can be easily desolved to
\begin{equation}
P(k)=2^{k-1} (k-1)! .
\end{equation}

\subsubsection{Number of true Lin-$k$-Opts}
However, this overall set of
possibilities contains many variants in which old edges
which were cut are reused in the new configuration, such that
the move is not a true Lin-$k$-Opt. Thus, in order to get
the number of true Lin-$k$-Opts, those variants have to be
subtracted from the overall number. As there are ${k \choose i}$
possibilities to choose $i$
old edges for the new tour if there were overall $k$ deleted edges,
the number of true
Lin-$k$-Opts is given by the recursive formula
\begin{equation}
T(k)=2^{k-1} (k-1)! -\sum_{i=0}^{k-1} {k \choose i} T(i) \,
\mbox{with} \, T(0)=1.
\end{equation}
The starting point of this recursion is $T(0)=1$, as there
is one possibility for the Lin-0-Opt, the identity move, in which no edge
is changed. Table \ref{tabelle} gives an overview of the numbers
of true Lin-$k$-Opts for small $k$, in the case that each partial sequence
contains at least two nodes and that the TSP is symmetric. We find
that there is one Lin-0-Opt, the identity move, no Lin-1-Opt,
as by cutting only one edge no new tour can be formed, one
Lin-2-Opt, four Lin-3-Opts, and so on.
In the case of an asymmetric TSP, each number here
has to be multiplied by a factor of 2.

\begin{table}
\caption{Number of true Lin-$k$-Opts if the TSP is symmetric
and if every partial sequence contains at least 2 nodes
\label{tabelle}}
\begin{tabular}{|r|r|}
\hline
$k$ & $T(k)$ \cr
\hline
0 & 1 \cr
1 & 0 \cr
2 & 1 \cr
3 & 4 \cr
4 & 25 \cr
5 & 208 \cr
6 & 2121 \cr
\hline
\end{tabular}
\end{table}

\subsection{General case}
\subsubsection{Number of overall possibilities}
In the general case, not every partial sequence contains
at least two nodes. There might be sequences containing only
one node which are surrounded by two sequences containing
more than one node in the old tour. Furthermore, there might be tuples of
neighboring sequences containing only one node each which
are surrounded by two sequences containing more than one
node, and so on.

Let $\alpha_0$ be the number of the partial sequences containing
more than one node, $\alpha_1$ be the number of sequences containing
only one node and surrounded by two sequences containing more
than one node in the old tour, $\alpha_2$ be the number of
tuples of one-node-sequences surrounded by two sequences containing more
than one node in the old tour, $\alpha_3$ be the number of
triples of one-node-sequences surrounded by two sequences containing more
than one node in the old tour, and so on. We will see that
$P$ and $T$ are no longer functions of $k$ only, but depend on
the entries of the vector $\vec{\alpha}=(\alpha_0,\alpha_1,\dots)$.

In the following, the assumption shall hold that not every
edge of the tour is cut, such that $k<N$ and $\alpha_0>0$.
Thus, one can always choose a partial sequence
consisting of two or more nodes as a starting point for
the creation of a new tour and fix its
direction, as we only consider the symmetric TSP here.

Starting with this fixed partial sequence, a new feasible tour
containing all nodes can be constructed by iteratively
selecting an other partial sequence and adding it
to the end of the growing partial tour. For the overall number
of possibilities for construcing a new tour, it plays no role
whether one-node-sequences were side by side in the old tour
or not. Thus, the number of possibilities $P(\vec{\alpha})$
is simply given as
\begin{equation}
P(\vec{\alpha})= P(\alpha_0,k-\alpha_0,0,\dots,0) .
\end{equation}
There are two
possible ways for adding a partial sequence containing
at least two nodes, but only one possibility for adding
a sequence with only one node. Thus, analogously to the
result above we get the result
\begin{equation}
P(\vec{\alpha}) = 2^{\alpha_0-1} (k-1)! .
\end{equation}
For the asymmetric TSP, this number has to be
multiplied with 2.

\subsubsection{Number of true Lin-$k$-Opts}
When calculating the number of true Lin-$k$-Opts, we need
to consider the spatial arrangement of the one-node-sequences
in the old tour. We have to distinguish between single
one-node-sequences, tuples, triples, quadruples, and so on,
i.e., we have to consider the $i$-tuples for each $i$
separately. Contrarily, we have no problems with the
spatial arrangement of partial sequences with at least
two nodes. In order to get the number of true Lin-$k$-Opts, we want to use
a trick by artificially blowing up partial sequences with only
one node to sequences with two nodes.
Let us first consider here that there are not only partial
sequences with at least two nodes but also isolated partial
sequences with only one node. (We thus first leave out the
tuples, triples, \dots of single-node-sequences in our
considerations, but it does not matter here whether there are
any such structures.)
We extend one one-node sequence to two nodes by doubling the node.
Thus, one gets $T(\alpha_0+1,\alpha_1-1,\dots)$ possibilities
for performing a true Lin-$k$-Opt instead of
$T(\alpha_0,\alpha_1,\dots)$ possibilities. By changing the
direction of this blown-up sequence, one can connect it
-- in contrast to before as it consisted of only one node --
to those nodes of the neighboring parts to which it was
connected before. There are two possibilities to connect it
this way to one of the two neighboring partial sequences
and one possibility to connect it this way to both of them.
But these cases are forbidden, such that we have to subtract
the number of these possibilities in which they get
connected and we achieve
the recursive formula
\begin{equation}
\begin{array}{lll} \displaystyle
T(\vec{\alpha})=\frac12 ( & & T(\alpha_0+1,\alpha_1-1,\alpha_2,\dots) \cr
\displaystyle & - 2 & T(\alpha_0,\alpha_1-1,\alpha_2,\dots) \cr
\displaystyle & - & T(\alpha_0-1,\alpha_1-1,\alpha_2,\dots) .\cr
\end{array}
\end{equation}
Please note that the resulting number has to be divided by 2,
as a partial sequence with only one node cannot be inserted in two different
directions.

Analogously, one can derive a formula if there are tuples of neighboring
sequences with only one node each. Here one expands one of the two
partial sequences to two nodes, such that there is one tuple less,
but one isolated one-node-sequence more and one longer sequence more.
Analogously to above, the false possibilities must be subtracted
and the result divided by 2, such that we get the formula
\begin{equation}
\begin{array}{lll} \displaystyle
T(\vec\alpha) =
\frac12 ( & & T(\alpha_0+1,\alpha_1+1,\alpha_2-1,\alpha_3,\dots) \cr
\displaystyle & - & T(\alpha_0,\alpha_1+1,\alpha_2-1,\alpha_3,\dots) \cr
\displaystyle & - & T(\alpha_0+1,\alpha_1,\alpha_2-1,\alpha_3,\dots) \cr
\displaystyle & - & T(\alpha_0,\alpha_1,\alpha_2-1,\alpha_3,\dots) ) . \cr
\end{array}
\end{equation}

For all longer groups of single-node-sequences, like
triples and quadruples, there is one common approach.
Here it is appropriate to blow up a single-node-sequence at
the frontier, such that the following recursive formula
is achieved:
\begin{equation}
\begin{array}{llll} \displaystyle
T(\vec\alpha) =
\displaystyle \frac12 ( & & T( & \alpha_0+1,\alpha_1,\dots,
  \alpha_{i-1}+1,\alpha_i-1, \cr
  & & & \alpha_{i+1},\dots) \cr
\displaystyle & - & T( & \alpha_0,\alpha_1,\dots,
  \alpha_{i-1}+1,\alpha_i-1, \cr
  & & & \alpha_{i+1},\dots) \cr
\displaystyle & - & T( & \alpha_0+1,\alpha_1,\dots,
  \alpha_{i-2}+1,\alpha_{i-1},\alpha_i-1, \cr
  & & & \alpha_{i+1},\dots) \cr
\displaystyle & - & T( & \alpha_0,\alpha_1,\dots,
  \alpha_{i-2}+1,\alpha_{i-1},\alpha_i-1, \cr
  & & & \alpha_{i+1},\dots) ) \cr
\end{array}
\end{equation}

Generally, one should proceed with the recursion in the following
way: first, those non-zero $\alpha_i$ should vanish for which $i$ is maximal.
This approach should be iterated with decreasing $i$ until
one ends up for a formula for tours with partial sequences consisting of
at least two nodes each
for which we can use the formula for the special case.

\section{Number of possibilities for cutting the tour
with a Lin-$k$-Opt\label{trennen}}
\subsection{Special Case}
\begin{table}
\caption{Number of possibilities for cutting the tour of a
traveling salesman if each partial sequence shall contain
at least two nodes: $k$ denotes the number of cuts of the
Lin-$k$-Opt, $C(k)$ the number of possibilities.
\label{sonderfall}}
\begin{tabular}{|l|l|}
\hline
$k$ & $C(k)$ \cr
\hline
2 & $N \times (N-3)/2$ \cr
3 & $N \times (N-4)(N-5)/3!$ \cr
4 & $N \times (N-5)(N-6)(N-7)/4!$ \cr
5 & $N \times (N-6)(N-7)(N-8)(N-9)/5!$ \cr
\hline
\end{tabular}
\end{table}

After having determined the number of possibilities to reconnect
partial sequences to a closed tour with a true Lin-$k$-Opt,
we still have to determine the number of possibilities for cutting
the tour in order to create these partial sequences.
We again start out with the special case in which every partial
sequence to be created shall contain at least two nodes.
By empirical going through all possibilities, we found the
formulas which are given in Tab.\ \ref{sonderfall}. From this
result, we deduce a general formula for the possibilities
$C(k)$ for the Lin-$k$-Opt:
\begin{equation} \label{formelspeziell}
C(k)=N \times \prod_{i=1}^{k-1} (N-k-i) \times
\frac{\displaystyle 1}{\displaystyle k!} =
\frac{\displaystyle N}{\displaystyle N-k} \times
{N-k \choose k}
\end{equation}
Thus, we find here that the neighborhood created by a Lin-$k$-Opt
is of order ${\cal O}(N^k)$.

\subsection{General Case}
\begin{table}
\caption{Number of possibilities for cutting the tour of a
traveling salesman: $k$ denotes the overall number of cuts
performed by the Lin-$k$-Opt, $\beta_1$, $\beta_2$, $\beta_3$,
and $\beta_4$ denote the number of $1$-, $2$-, $3$-, and $4$-type
multicuts as defined in the text. Only $\beta_i$-values for
$i\le 4$ are considered here in our examples. For the Lin-6-Opt,
only some special cases are considered.
The number $C$ of possibilities depends on all these numbers.
All these formulas for $C(\vec{\beta})$ were found by hand.
\label{allgemeinerfall}}
\begin{tabular}{|l|llll|l|}
\hline
$k$ & $\beta_1$ & $\beta_2$ & $\beta_3$ & $\beta_4$ & $C$ \cr
\hline
2 & 2 & 0 & 0 & 0 & $N \times (N-3)/2$ \cr
  & 0 & 1 & 0 & 0 & $N$ \cr
\hline
3 & 3 & 0 & 0 & 0 & $N \times (N-4)(N-5)/3!$ \cr
  & 1 & 1 & 0 & 0 & $N \times (N-4)$ \cr
  & 0 & 0 & 1 & 0 & $N$ \cr
\hline
4 & 4 & 0 & 0 & 0 & $N \times (N-5)(N-6)(N-7)/4!$ \cr
  & 2 & 1 & 0 & 0 & $N \times (N-5)(N-6)/2$ \cr
  & 0 & 2 & 0 & 0 & $N \times (N-5)/2$ \cr
  & 1 & 0 & 1 & 0 & $N \times (N-5)$ \cr
  & 0 & 0 & 0 & 1 & $N$ \cr
\hline
5 & 5 & 0 & 0 & 0 & $N \times (N-6)(N-7)(N-8)(N-9)/5!$ \cr
  & 3 & 1 & 0 & 0 & $N \times (N-6)(N-7)(N-8)/3!$ \cr
  & 1 & 2 & 0 & 0 & $N \times (N-6)(N-7)/2$ \cr
  & 2 & 0 & 1 & 0 & $N \times (N-6)(N-7)/2$ \cr
  & 0 & 1 & 1 & 0 & $N \times (N-6)$ \cr
  & 1 & 0 & 0 & 1 & $N \times (N-6)$ \cr
\hline
6 & 2 & 2 & 0 & 0 & $N \times (N-7)(N-8)(N-9)/2/2$ \cr
  & 0 & 3 & 0 & 0 & $N \times (N-7)(N-8)/3!$ \cr
  & 3 & 0 & 1 & 0 & $N \times (N-7)(N-8)(N-9)/3!$ \cr
  & 1 & 1 & 1 & 0 & $N \times (N-7)(N-8)$ \cr
  & 0 & 0 & 2 & 0 & $N \times (N-7)/2$ \cr
  & 2 & 0 & 0 & 1 & $N \times (N-7)(N-8)/2$ \cr
  & 0 & 1 & 0 & 1 & $N \times (N-7)$ \cr
\hline
\end{tabular}
\end{table}

In the general case, an arbitrary Lin-$k$-Opt can also lead to
partial sequences containing only one node. Here we have
to distinguish between various types of cuts: The cuts introduced
by a Lin-$k$-Opt can be isolated, i.e., they are between two
sequences with more than one node each. Then they can lead to
isolated nodes which are between two partial sequences with
more than one node each, and so on.
Let us view this from the point of view of the cuts of the
tour. All in all, a Lin-$k$-Opt generally leads to $k$
cuts in the tour. Let us denote an $i$-type multicut (with
$1 \le i \le k$) at position $j$ (with $1 \le j \le N$)
as the scenario that the tour is cut at $i$ successive
positions after the node with the tour position number $j$.
Thus, the tour is cut by an $i$-type multicut
successively between pairs of nodes
with the tour position numbers $(j,j+1),(j+1,j+2),
\dots,(j+i-1,j+i)$.

Furthermore, let $\beta_i$ be the number of $i$-type multicuts:
$\beta_1$ is the number of isolated cuts.
$\beta_2$ is the number of 2-type multicuts by which the tour
is cut after two successive tour positions such that
a partial sequence containing only one node is created,
surrounded by two sequences containing more than one node.
Thus, $\beta_2$ is also the number of isolated nodes
surrounded by longer sequences and is thus identical
with $\alpha_1$. Analogously, 3-type multicuts
lead to tuples of nodes which are surrounded by partial
sequences with more than one node, thus, $\beta_3$ is
the number $\alpha_2$ of these tuples. Analogously, 4-type multicuts
lead to triples of sequences containing only one node each,
and so on. Generally, we have for all $i\ge 1$ that
\begin{equation}
\alpha_i=\beta_{i+1}
\end{equation}
of the last section, but for $i=0$ the situation is different:
each $i$-type multicut produces a further sequence consisting of
at least two nodes, such that we have
\begin{equation}
\alpha_0 = \sum_{i=1}^k \beta_i.
\end{equation}
Please note that $\alpha_0$ is both the number of these longer
partial sequences and the number of all $i$-type multicuts.
The overall number $k$ of cuts can be expressed as
\begin{equation}
k = \sum_{i=1}^k i\beta_i.
\end{equation}

In this general case, the number $C$ of ways of cutting the
tour also depends not only on $k$, but on the entries
of the vector $\vec{\beta}=(\beta_1,\beta_2,\dots,\beta_k)$.
As Tab.\ \ref{allgemeinerfall} shows, the order of the neighborhood
size is now given as ${\cal O}(N^{\alpha_0})$. From these
examples in the table, we empirically derive the formula
\begin{equation}
C(\vec{\beta})=N \times \prod_{i=1}^{\sum_{j=1}^k \beta_j-1} (N-k-i) \times
\frac{\displaystyle 1}{\displaystyle \prod_{i=1}^k \beta_i!}
\end{equation}
for the number of possibilities for cutting a tour with
a Lin-$k$-Opt, leading to $\beta_i$ many $i$-type multicuts.
Please note that if the upper index of a product is smaller
than the lower index, then this so-called empty product is 1.
This formula can be rewritten to
\begin{equation} \label{formelallgemein}
C(\vec{\beta})=\frac{\displaystyle N}{\displaystyle N-k} \times
{N-k \choose \alpha_0} \times
\frac{\displaystyle \alpha_0!}{\displaystyle \prod_{i=1}^k \beta_i!}
\end{equation}
making use of $\alpha_0$ being the sum of all $\beta_i$.
Please note that Eq.\ (\ref{formelspeziell}) for the special
case with each sequence containing at least two nodes is
a special case of Eq.\ (\ref{formelallgemein}).

The numbers for cutting a tour in partial sequences in this
section were first found by hand for the examples given in
Tabs.\ \ref{sonderfall} and \ref{allgemeinerfall},
then the general formulas were intuitively deduced from
these. After that the correctness of these fomulas
was checked by computer programs up to $k=20$
for the special case and for all variations of the general
case.

\section{Quality of the Results Achieved with Various Moves}

\begin{table} \centering
\caption{Results for the Greedy Algorithm using small moves
(Exchange, Node Insertion Move, and Lin-2-Opt) and the
variants of the Lin-3-Opt: for each
instance, 100 optimization runs were performed, starting with a random
configuration and performing a specific number (given in the text)
of the corresponding move.}
\label{ergebnisse}
\begin{tabular}{|l|c|r|r|rcl|}
\hline
instance& move & minimum & maximum & mean value & $\pm$ & error \\
\noalign{\smallskip}\hline\noalign{\smallskip}
BEER127 & EXC & 159705.087 & 208409.703 & 182672.403 & $\pm$ & 977.4 \\
        & NIM & 132898.075 & 161648.200 & 146418.726 & $\pm$ & 596.4 \\
        & L2O & 121178.315 & 138126.861 & 129964.407 & $\pm$ & 325.5 \\
  & all small & 119331.431 & 134236.701 & 126605.995 & $\pm$ & 305.2 \\
        & L3O1 & 119218.371 & 131235.436 & 125118.494 & $\pm$ & 254.3 \\
        & L3O2 & 120935.053 & 133226.981 & 125888.805 & $\pm$ & 260.4 \\
        & L3O3 & 118589.344 & 127980.167 & 121981.992 & $\pm$ & 187.4 \\
        & L3O4 & 118899.537 & 127588.470 & 121928.533 & $\pm$ & 187.4 \\
        &L3Oall& 118629.043 & 127881.507 & 121396.175 & $\pm$ & 194.1 \\
\hline
LIN318  & EXC & 90116.6104 & 127763.120 & 110660.750 & $\pm$ & 628.5 \\
        & NIM & 59992.7967 & 78436.1451 & 68407.9569 & $\pm$ & 376.2 \\
        & L2O & 45115.9243 & 49971.4471 & 47276.4710 & $\pm$ & 95.0 \\
  & all small & 43488.2914 & 47875.0510 & 45743.4573 & $\pm$ & 78.9 \\
        & L3O1 & 43883.1117 & 46992.0997 & 45210.0271 & $\pm$ & 63.9 \\
        & L3O2 & 44209.8503 & 47368.0175 & 45466.4980 & $\pm$ & 66.1 \\
        & L3O3 & 42863.7951 & 44822.3800 & 43632.8142 & $\pm$ & 39.5 \\
        & L3O4 & 42626.6703 & 45173.6630 & 43534.8133 & $\pm$ & 44.4 \\
        &L3Oall& 42401.9352 & 44514.0820 & 43518.6735 & $\pm$ & 40.7 \\
\hline
PCB442  & EXC & 112322.878 & 142974.558 & 129290.471 & $\pm$ & 708.4 \\
        & NIM & 64129.9159 & 79939.0300 & 70764.4069 & $\pm$ & 296.7 \\
        & L2O & 54682.4205 & 59026.1119 & 56614.6608 & $\pm$ & 82.3 \\
  & all small & 52589.4816 & 56840.8241 & 54660.7005 & $\pm$ & 76.3 \\
        & L3O1 & 53547.9002 & 56348.0251 & 54850.7759 & $\pm$ & 60.7 \\
        & L3O2 & 52975.6709 & 57838.1568 & 54847.6013 & $\pm$ & 85.7 \\
        & L3O3 & 51658.9173 & 54163.2202 & 52689.2427 & $\pm$ & 47.9 \\
        & L3O4 & 51627.2806 & 53519.6159 & 52545.4050 & $\pm$ & 41.0 \\
        &L3Oall& 51480.2480 & 53892.2666 & 52493.4278 & $\pm$ & 44.5 \\
\hline
ATT532  & EXC & 69294 & 91946 & 80037.86 & $\pm$ & 449.7 \\
        & NIM & 37691 & 47862 & 42460.43 & $\pm$ & 212.1 \\
        & L2O & 29960 & 32026 & 30867.73 & $\pm$ & 42.2 \\
  & all small & 29069 & 30530 & 29689.64 & $\pm$ & 31.1 \\
        & L3O1 & 28716 & 30639 & 29732.22 & $\pm$ & 33.7 \\
        & L3O2 & 29229 & 30817 & 30031.61 & $\pm$ & 30.7 \\
        & L3O3 & 28275 & 29369 & 28692.05 & $\pm$ & 22.8 \\
        & L3O4 & 28076 & 29159 & 28673.47 & $\pm$ & 21.4 \\
        &L3Oall& 28281 & 29102 & 28718.27 & $\pm$ & 19.6 \\
\hline
NRW1379 & EXC & 164670.848 & 196316.798 & 181017.337 & $\pm$ & 607.0 \\
        & NIM & 89353.7705 & 105344.742 & 94486.5234 & $\pm$ & 294.3 \\
        & L2O & 62862.6385 & 64887.1349 & 64046.8071 & $\pm$ & 42.9 \\
  & all small & 60334.7758 & 62208.5391 & 61154.0388 & $\pm$ & 39.4 \\
        & L3O1 & 64624.4812 & 66712.2867 & 65611.2830 & $\pm$ & 35.6 \\
        & L3O2 & 63267.3838 & 64788.3074 & 64008.2578 & $\pm$ & 37.1 \\
        & L3O3 & 58730.9852 & 59768.7048 & 59199.8516 & $\pm$ & 25.4 \\
        & L3O4 & 58727.9973 & 59656.5893 & 59173.3947 & $\pm$ & 23.5 \\
        &L3Oall& 58650.2000 & 60077.9761 & 59307.8550 & $\pm$ & 26.1 \\
\hline
\end{tabular}
\end{table}

Finally, we want to ask which quality of solutions can be
achieved with the various moves. Here we concentrate on
a comparison of the quality which can be achieved by the
``power'' of the moves only, thus, we do not use some
elaborate underlying
heuristics like those mentioned in the introduction, bnt
use the simplest algorithm, which is often called Greedy Algorithm, here:
starting out from a randomly created configuration
$\sigma_0$, a series
of moves is applied to the system. Each move chooses randomly
a way how to change the current configuration $\sigma_i$ into
a new configuration $\sigma_{i+1}$. A move is only accepted
by the Greedy Algorithm
if it does not lead to a deterioration, i.e., if the length
${\cal H}(\sigma_{i+1})$
of the tentative new tour is as long as or is shorter than
the length ${\cal H}(\sigma_i)$ of the current tour. After the
acceptance of the move $\sigma_i\to\sigma_{i+1}$, the system
tries to move from the configuration $\sigma_{i+1}$ to a
new configuration.
In case of rejection of the move $\sigma_i\to\sigma_{i+1}$,
one sets $\sigma_{i+1}=\sigma_i$ and proceeds as above.

Table \ref{ergebnisse} shows the minimum, maximum, and
average quality of solutions which can be achieved with
the various moves for five different TSP benchmark instances
which were taken from Reinelt's TSPLIB95 \cite{TSPLIB95}
and which vary in size between $N=127$ and $N=1379$.
In order to be quite sure to end up
in a local minimum, we used $50\times N^2$ move trials if using
either the Exchange or the Node Insertion Move or the Lin-2-Opt.
If comparing the results achieved with these small moves, we find
that the Lin-2-Opt leads to the best results and the Exchange
leads to the worst results, a result which is in accordance
with the theoretical derivations in \cite{Stadler}.
Now one can also mix these moves in
the way that a general move routine calls each of these three
moves with equal probability. The next line in Tab.\
\ref{ergebnisse} shows that the results, which were taken
after $150\times N^2$ move trials, are better for this mixture
than for any of the three moves themselves. This is of course
due to the larger neighborhood size of this mixture.

Then we provide the results for the four variants of the
Lin-3-Opt, for which the results were taken after
$10\times N^3$ move trials (A few test runs froze after $1-2 \times N^3$
move trials),
and for a mixture (Here we call each of the four variants
with equal probability), for which the results were taken
after $20\times N^3$ move trials. If we denote the tour position
numbers after which the tour is cut by the Lin-3-Opt as $i$,
$j$, and $k$ with $i<j<k$ and their successive tour position
numbers as $i_+$, $j_+$, and $k_+$, we can write the cut tour as
follows:
$$
\begin{array}{cc|ccc|ccc|cc}
\dots & \sigma(i) & \sigma(i_+) &
\dots & \sigma(j) & \sigma(j_+) &
\dots & \sigma(k) & \sigma(k_+) & \dots
\end{array}
$$
Please note that the tour is of course closed, such that
the partial sequence starting at $\sigma(k_+)$ ends with
$\sigma(i)$. If
$j \neq i_+$, $k \neq j_+$, and $i \neq k_+$,
there are four possibilities to reconnect the
three partial sequences with a true Lin-3-Opt:
$$
\begin{array}{ccc|ccc|ccc|cc}
\mbox{L3O1:} & \dots & \sigma(i) & \sigma(j_+) &
\dots & \sigma(k) & \sigma(i_+) &
\dots & \sigma(j) & \sigma(k_+) & \dots \cr
\mbox{L3O2:} & \dots & \sigma(i) & \sigma(j) &
\dots & \sigma(i_+) & \sigma(k) &
\dots & \sigma(j_+) & \sigma(k_+) & \dots \cr
\mbox{L3O3:} & \dots & \sigma(i) & \sigma(j_+) &
\dots & \sigma(k) & \sigma(j) &
\dots & \sigma(i_+) & \sigma(k_+) & \dots \cr
\mbox{L3O4:} & \dots & \sigma(i) & \sigma(k) &
\dots & \sigma(j_+) & \sigma(i_+) &
\dots & \sigma(j) & \sigma(k_+) & \dots \cr
\end{array}
$$
Table \ref{ergebnisse} shows that there are
some differences in the qualities of the results achieved with the
four variants of the Lin-3-Opt and that the mixture provides
the best results for the three smaller instances, whereas two variants
are on average better than the mixture for the instances with $N=532$
and $N=1379$. Comparing these results with the results for the
small moves, we find that the four variants of the Lin-3-Opt
lead to a much better quality of the results than the small moves,
which is in accordance with the results in \cite{3OptBesser}.
As Stadler and Schnabl already mentioned in \cite{Stadler},
this better quality of the results has to be expected because
of the much larger neighborhood size of the Lin-3-Opts.

\begin{table} \centering
\caption{Results for the Greedy Algorithm using two types of Lin-4-Opts:
for each
instance, 100 optimization runs were performed, starting with a random
configuration and performing $N^4$ times one variant of the
Lin-4-Opt with the Greedy acceptance criterion. As the computing time
is rather large, only small instances are considered.}
\label{l4oergebnisse}
\begin{tabular}{|l|c|r|r|rcl|}
\hline
instance& move & minimum & maximum & mean value & $\pm$ & error \\
\hline
BEER127 & L4O1 & 123767.073 & 134986.024 & 129588.275 & $\pm$ & 244.2 \\
        & L4O2 & 121407.255 & 131300.030 & 125622.863 & $\pm$ & 214.7 \\
\hline
LIN318  & L4O1 & 45184.3246 & 48569.5839 & 46740.9896 & $\pm$ & 72.2 \\
        & L4O2 & 44539.2015 & 46932.1063 & 45451.7793 & $\pm$ & 48.3 \\
\hline
PCB442  & L4O1 & 55673.1534 & 58855.2662 & 57381.3213 & $\pm$ & 71.3 \\
        & L4O2 & 53793.3942 & 56843.4938 & 55209.2331 & $\pm$ & 64.0 \\
\hline
\end{tabular}
\end{table}

Now one can ask why not to proceed and to move on to even larger
moves. We implemented two of the 25 different variants of the Lin-4-Opt.
If denoting the tour position numbers after which the which the tour
is cut as $i$, $j$, $k$, and $l$ and their successive numbers as
$i_+$, $j_+$, $k_+$, and $l_+$, then the cut tour can be written as
follows:
$$
\begin{array}{cc|ccc|ccc|ccc|cc}
\dots & \sigma(i) & \sigma(i_+) &
\dots & \sigma(j) & \sigma(j_+) &
\dots & \sigma(k) & \sigma(k_+) &
\dots & \sigma(l) & \sigma(l_+) & \dots
\end{array}
$$
Then the move variants lead to the following new tours:
$$
\begin{array}{ccc|ccc|ccc|ccc|cc}
\mbox{L4O1:} & \dots & \sigma(i) & \sigma(k_+) &
\dots & \sigma(l) & \sigma(j_+) &
\dots & \sigma(k) & \sigma(i_+) &
\dots & \sigma(j) & \sigma(l_+) & \dots \cr
\mbox{L4O2:} & \dots & \sigma(i) & \sigma(k_+) &
\dots & \sigma(l) & \sigma(k) &
\dots & \sigma(j_+) & \sigma(i_+) &
\dots & \sigma(j) & \sigma(l_+) & \dots \cr
\end{array}
$$
The variant L4O1 is sometimes called the two-bridge-move,
as the two ``bridges'' $\sigma(i_+)\dots\sigma(j)$ and
$\sigma(k_+)\dots\sigma(l)$ are exchanged.
Table \ref{l4oergebnisse} shows the quality of the results
achieved with these two variants of the Lin-4-Opt.
We find that a further improvement cannot be found,
the results for the better variant of the Lin-4-Opt are
roughly of the same quality as the results for the
worse variants of the Lin-3-Opt. Thus, leaving the Local
Search approach even further leads to worse results.

Of course, using only the Greedy Algorithm, one fails
in achieving the global optimum configurations for the
TSP instances, which have a length of
118293.52\dots (BEER127 instance),
42042.535\dots (LIN318 instance), and
50783.5475\dots (PCB442 instance),
respectively. These optima
can be achieved with the small moves and the
variants of the Lin-3-Opt
if using a better underlying heuristic
(see e.g.\ \cite{sfb,sfb2,Bouncing}).

.

\section{Summary}
For getting an approximate solution of an instance
of the Traveling Salesman Problem, mostly an improvement
heuristic is used, which applies a sequence of move trials
which are either accepted or rejected according to the
acceptance criterion of the heuristics.
For the Traveling Salesman Problem, mostly small moves
are used which do not change the configuration very much.
Among these moves, the Lin-2-Opt, which cuts two edges of the
tour and turns around a
part of the tour, has been proved to provide superior results.
Thus, this Lin-2-Opt and its higher-order variants, the Lin-$k$-Opts,
which cut $k$ edges of the tour and reconnect the created
partial sequences to a new feasible solution, and their
properties have drawn great attention.

In this paper,
we have provided formulas for the exact calculation of the number
of configurations which can be reached from an arbitrarily
chosen tour via these Lin-$k$-Opts. A specific Lin-$k$-Opt
leads to a certain structure of multicuts, i.e., there are
isolated cuts, which divide two partial sequences with at
least two nodes, then there are two cuts just behind each
other, such that a partial sequence with only one node
is created, which is in between two partial sequences with
more than one node, then there are three cuts just behind
each other, such that a tuple of partial sequences with
only one node each is created, and so on. The number of
possibilities for cutting a tour according to these
structures of multicuts is given in Eq.\ (14).
From the numbers of multicut structures, one has then to 
derive the numbers of partial sequences, which are given
in Eqs.\ (10) and (11). Then one has to use the recursive formulas
(6-8) in order to simplify the dependency of the number of
reconnections to only one parameter. Finally, one has to use
Eq.\ (3) for getting the number of reconnections for a true
Lin-$k$-Opt with $k$ new edges. Finally, one has to sum
up the products of the numbers of possible cuttings and of the
numbers of possible reconnections in order to get the
overall number of neighboring configurations which can
be reached via the move.

At the end, we have compared the results achieved with
these moves using the simple Greedy Algorithm which
rejects all moves leading to deteriorations.
We have found that the Lin-2-Opt is superior to the other
small moves and that the Lin-3-Opt provides even
better results than the Lin-2-Opt.
But moving even further away from the Local Search approach
does not lead to further improvements.

\begin{acknowledgments}
J.J.S. wants to thank Erich P. Stoll (University of Zurich)
for the picture of the United States of America, which serves
as a background in Fig.\ \ref{fig:usa}.
\end{acknowledgments}

\end{document}